\documentclass[12pt]{article}
\begin{document}

\title{Anomaly of Dark Matter}
\author{B.G. Sidharth\footnote{birlasc@gmail.com}, B.M. Birla Science Centre,\\ Adarsh Nagar, Hyderabad - 500 063, India\\
\\Abhishek Das\footnote{parbihtih3@gmail.com}\footnote{$^*$ Invited talk at Physics and Astronomy Conference 2018,
Hyderabad, India}, B.M. Birla Science Centre,\\ Adarsh Nagar,
Hyderabad - 500 063, India}
\date{}
\maketitle

\begin{abstract}
Latest observations by Riess and coworkers in 2017 have reconfirmed
their earlier observation that the universe is accelerating some
eight percent faster than the currently accepted cosmological model
described. In this paper we argue that this discrepancy can be
eliminated by considering a universe consisting only of matter and
dark energy. Interestingly, recently Mark Swissbank, R. Genzel et
al. have concluded that nearly
ten billion years ago dark matter concentration was negligible.\\ \\

{\bf keywords}: cosmology: dark energy - cosmology: dark matter
\end{abstract}

\maketitle

\section{Introduction}
Hubble's law is regarded as one of the major observational basis for
the expansion of the universe. Later the existence of {\it dark
matter} was hypothesized by Zwicky \cite{Zwicky1,Zwicky2} who
inferred the existence some unseen matter based on his observations
regarding the rotational velocity curves at the edge of galaxies.
Although, Jacobus Kapteyn \cite{Kapteyn} and Jan Oort had
\cite{Oort} also had derived the same conclusions before Zwicky.
Since then, various efforts have been made to prove the existence of
{\it dark matter} [cf.ref. \cite{Bertone} for detailed review].
Recently, after conducting experiments to detect weakly interacting
massive particles (WIMPs) that interact only through gravity and the
weak force and are hypothesized as the constituents of {\it dark matter} that have led nowhere \cite{Akerib}.\\
Interestingly, authors such as Milgrom \cite{mil1}, Bekenstein
\cite{Bek} Sidharth (Cf. Section 3) and Mannheim \cite{Mann} have
endeavoured to find alternatives to the widely accepted {\it dark
matter}. The author Sidharth \cite{bgs1,bgs2,bgs3} has also given a
suitable alternative to the conventional {\it dark matter} paradigm.
Nevertheless, the objective of this paper is to substantiate that
the existence of {\it dark matter} is inconsistent with the recent
observations made by Riess \cite{Riess} regarding the Hubble's
constant.\\
The generally accepted ideas may have to be revisited in view of
latest observations of Riess et al which point to the fact that
cosmic acceleration is some $5-8 \%$ greater than what the current
cosmological model suggests.\\
Before proceeding, it may be mentioned that in 1997, the accepted
model of the universe was that of a dark matter dominated
decelerating universe. That year the author Sidharth put forward his
contra model -- an accelerating universe, dominated by not dark
matter but rather what is today being called dark energy. As Tony
Leggett put it,"... It is of course clear that your equation
predicts an exponential (inflation-type) expansion of the current
universe, hence acceleration. And it would have been nice if the
Nobel committee had mentioned this, ..." and "... I certainly do
appreciate that you are one of the very few to have recognized, on
theoretical grounds, the possible need to reintroduce a nonzero
cosmological constant ahead of the supernova experiment!"
\cite{leggett}.

\section{Theory}

We are well acquainted with the fact that the Friedman equations govern the expansion of space in homogeneous and isotropic models of the universe
within the context of general relativity. Let us begin with the following equation\\
\[H^{2} = (\frac{\dot{a}}{a})^{2} = \frac{8\pi G}{3}\rho - \frac{kc^{2}}{a^{2}} + \frac{\Lambda c^{2}}{3}\]\\
where $H$ is the Hubble parameter, $a$ is the scale factor, $G$ is the gravitational constant,
$k$ is the normalized spatial curvature of the universe and $\Lambda$ is the cosmological constant.
Considering $k = 0$ (a flat universe) with the domination of both matter and dark energy, one can
derive the Hubble parameter as\\
\begin{equation}
H(z) = H_{0}[\Omega_{M}(1 + z)^{3} + \Omega_{DE}(1 + z)^{3(1 + w)}]^{\frac{1}{2}} \label{a}
\end{equation}
where, $z$ is the redshift value or the recessional velocity and the dimensionless parameter $w$ is given by\\
\[P = w\rho c^{2}\]\\
$P$ being the pressure and $\rho$ being the density. Now, we would like to expand the function $H(z)$ using the Taylor expansion about the point $z_{0}$.
This yields\\
\[H(z) = H(z_{0}) + \frac{H^{\prime}(z_{0})}{1!} (z - z_{0}) + \cdots\]\\
Neglecting terms consisting second and higher order derivatives of the Hubble parameter and considering that $H(z_{0}) = H_{0}$ we have using (\ref{a})\\
\begin{equation}
H(z) = H_{0} + \frac{H_{0}}{2} \frac{3\Omega_{M}(1 + z_{0})^{2} + 3(1 + w)\Omega_{DE}(1 + z_{0})^{3(1 + w) - 1}}{[\Omega_{M}(1 + z_{0})^{3}
+ \Omega_{DE}(1 + z_{0})^{3(1 + w)}]^{\frac{1}{2}}} (z - z_{0})
\end{equation}
Now, we know that if the dark energy derives from a cosmological constant then \\
\[w = -1\]\\
Therefore, in such a case we have\\
\begin{equation}
H(z) = H_{0} + \frac{3H_{0}}{2} \frac{\Omega_{M}(1 + z_{0})^{2}}{[\Omega_{M}(1 + z_{0})^{3} + \Omega_{DE}]^{\frac{1}{2}}} (z - z_{0})
\end{equation}
Now, since numerical values suggest that $\Omega_{DE} > \Omega_{M}$ we can use another series expansion for the denominator of the second term above to get\\
\[H(z) = H_{0} + \frac{3H}{2}\frac{1}{\sqrt{\Omega_{DE}}}[\Omega_{M}(1 + z_{0})^{2}][1 - \frac{\Omega_{M}(1 + z_{0})^{2}}{2\Omega_{DE}}](z - z_{0})\]\\
Thus, we can write finally\\
\begin{equation}
H(z) = H_{0}[1 + \frac{3}{2}\frac{1}{\sqrt{\Omega_{DE}}}\{\Omega_{M}(1 + z_{0})^{2}\}\{1 - \frac{\Omega_{M}(1 + z_{0})^{2}}{2\Omega_{DE}}\}](z - z_{0})
\end{equation}
Now, we would look at this equation at the point $z_{0} = 0$ and for $z = 1$ to give\\
\begin{equation}
H = H_{0}[1 + \frac{3}{2}\frac{\Omega_{M}}{\sqrt{\Omega_{DE}}}\{1 - \frac{\Omega_{M}}{2\Omega_{DE}}\}] \label{b}
\end{equation}
Now, standard cosmological model suggests that the universe is comprised of baryonic matter, dark matter, dark energy and some other constituents. In a
nutshell, we have \cite{Knop}\\
\[\Omega_{Baryonic} \approx 0.04\]\\
\[\Omega_{Dark matter} \approx 0.23\]\\
\[\Omega_{Dark energy} \approx 0.73\]\\
and\\
\[\Omega_{M} = \Omega_{Baryonic} + \Omega_{Dark matter}\]
Using all these values in (\ref{b}) we have the Hubble parameter\\
\[H = H_{0} + 0.39H_{0}\]\\
i.e. the acceleration of the universe should be approximately $39\%$ greater than it's value. But, due to recent observations it has been
substantiated that the acceleration is about $5\%-8\%$ greater than it's value. So, in fact we should have\\
\[H = H_{0} + 0.08H_{0}\]\\
If this is the case then doing some back calculations and using $\Omega_{DE} \approx 0.73$, we arrive at a quadratic equation in $\Omega_{M}$ as\\
\begin{equation}
(1 - 0.685\Omega_{M})\Omega_{M} = 0.045
\end{equation}
Solving this equation we have the following two values for $\Omega_{M}$.\\
\[\Omega_{M} \approx 1.41 ~or, 0.044\]\\
Now, it is a fact that $\Omega_{M} < 1$ and $\Omega_{M}$ would be unphysical. Therefore we have the value of $\Omega_{M}$ as\\
\begin{equation}
\Omega_{M} \approx 0.044
\end{equation}
But, this is very nearly equal to the value of Baryonic matter, i.e. $\Omega_{Baryonic}$. This suggests ostensibly that \\
\begin{equation}
\Omega_{Dark matter} \approx 0
\end{equation}
In other words, the existence of dark matter is itself inconsistent according to the latest observations of Riess et al.
In such a case, the total density of the universe is given by\\
\begin{equation}
\Omega = \Omega_{Baryonic} + \Omega_{Dark energy} \approx 0.77
\end{equation}
which is less than the critical density. This suggests that the
universe will be expanding in an accelerating manner.\\

\section{Alternative to the {\it dark matter} paradigm}

Very recently the LUX detector in South Dakota has concluded \cite{Akerib} that it has not found any traces of Dark Matter. So far
this has been the most delicate detector. It will be recalled that
dark matter was introduced in the 1930s by F. Zwicky to explain the
flattening of the galactic rotational curves: With Newtonian gravity
the speeds of these galactic curves at the edges should tend to zero
according to the Keplerian law, $v \propto 1/\sqrt{r}$. Here $r$ is
the distance to the edge from the galactic centre. However velocity
$v$ remains more or less constant. Zwicky explained this by saying
that there is a lot more of unseen matters concealed in the
galaxies, causing this
discrepancy. The fact is that even after nearly 90 years dark matter has not been detected.\\
The modified Newtonian dynamics approach of Milgrom
\cite{mil1,mil2,mil3,mil4,mil5,milgrom} was an interesting
alternative to the {\it dark matter} paradigm. The objection of this
fix has been that it is too ad hoc, without any
underlying theory.\\
The author himself has been arguing over the years
\cite{bgs1,bgs2,bgs3} (Cf.ref.\cite{tduniv} for a summary) that the
gravitational constant $G$ is not fixed but varies slowly with time
in a specific way. In fact this variation of the gravitational
constant has been postulated by Dirac, Hoyle and others from a
different point of view (Cf.ref.\cite{tduniv,narlikar}) which for
various reasons including inconsistencies have in the author's
scheme, exactly accounts for the galactic rotation anomaly without
resorting to dark matter or without contradictions.\\
Our starting point is the rather well known relation \cite{narlikar}
\begin{equation}
G = G_o (1- \frac{t}{t_o})\label{3ey15}
\end{equation}
where $G_o$ is the present value of $G$ and $t_o$ is the present age
of the Universe, while $t$ is the relatively small time elapsed from
the present epoch. On this basis one could correctly explain the
gravitational bending of light, the precession of the equinoxes of
mercury, the shortening of the orbits of binary pulsars and even the
anomalous acceleration of the pioneer spacecrafts (Cf.references
given above).\\
Returning to the problem of the rotational velocities at the edges
of galaxies, one would expect these to fall off according to
\begin{equation}
v^2 \approx \frac{GM}{r}\label{3ey33}
\end{equation}
However it is found that the velocities tend to a constant value,
\begin{equation}
v \sim 300km/sec\label{3ey34}
\end{equation}
This, as noted, has lead to the postulation of the as yet undetected
additional matter alluded to, the so called dark matter.(However for
an alternative view point Cf.\cite{sivaramfpl93}). We observe that
from (\ref{3ey15}) it can be easily deduced that\cite{cu,bgsedge}
\begin{equation}
a \equiv (\ddot{r}_{o} - \ddot{r}) \approx \frac{1}{t_o}
(t\ddot{r_o} + 2\dot r_o) \approx -2 \frac{r_o}{t^2_o}\label{3ey35}
\end{equation}
as we are considering infinitesimal intervals $t$ and nearly
circular orbits. Equation (\ref{3ey35}) shows
(Cf.ref\cite{bgs2} also) that there is an anomalous inward
acceleration, as if there is an extra attractive force, or an
additional central mass, a la Zwicky's dark matter.\\
So,
\begin{equation}
\frac{GMm}{r^2} + \frac{2mr}{t^2_o} \approx
\frac{mv^2}{r}\label{3ey36}
\end{equation}
From (\ref{3ey36}) it follows that
\begin{equation}
v \approx \left(\frac{2r^2}{t^2_o} + \frac{GM}{r}\right)^{1/2}
\label{3ey37}
\end{equation}
From (\ref{3ey37}) it is easily seen that at distances within the
edge of a typical galaxy, that is $r < 10^{23}cms$ the equation
(\ref{3ey33}) holds but as we reach the edge and beyond, that is for
$r \geq 10^{24}cms$ we have $v \sim 10^7 cms$ per second, in
agreement with (\ref{3ey34}). In fact as can be seen from
(\ref{3ey37}), the first term in the square root has an extra
contribution (due to the varying $G$) which is roughly some three to
four times the
second term, as if there is an extra mass, roughly that much more.\\
Thus the time variation of G explains observation without invoking
dark matter.
\section{Conclusion}
We have seen that the discrepancy in the acceleration value of the
universe, as reconfirmed multiple times by careful studies of Riess
and coworkers can be removed by considering a universe consisting
only of matter and
dark energy.\\
Even more recently, Mark Swissbank \cite{Mark} and R. Genzel et al. \cite{Genzel} have concluded that nearly ten billion years ago dark matter
concentration was very small and the universe was dominated by baryonic matter. It is possible that this negligible concentration of dark matter
boiled down to zero in due course of evolution of the universe.

\end{document}